\documentclass{JHEP3}
\usepackage{epsfig}
\usepackage{amssymb}
\usepackage{amscd}
\usepackage{amstext}
\usepackage{graphics}

\newcommand{\fref}[1]{Figure~\ref{#1}}
\newcommand{\cref}[1]{Chapter~\ref{#1}}
\newcommand{\beq}{\begin{equation}}
\newcommand{\eeq}{\end{equation}}
\newcommand{\ba}{\begin{array}}
\newcommand{\ea}{\end{array}}
\newcommand{\bcenter}{\begin{center}}
\newcommand{\ecenter}{\end{center}}

\def\IB{\relax\hbox{$\inbar\kern-.3em{\rm B}$}}
\def\IC{\relax\hbox{$\inbar\kern-.3em{\rm C}$}}
\def\ID{\relax\hbox{$\inbar\kern-.3em{\rm D}$}}
\def\IE{\relax\hbox{$\inbar\kern-.3em{\rm E}$}}
\def\IF{\relax\hbox{$\inbar\kern-.3em{\rm F}$}}
\def\IG{\relax\hbox{$\inbar\kern-.3em{\rm G}$}}
\def\IGa{\relax\hbox{${\rm I}\kern-.18em\Gamma$}}
\def\IH{\relax{\rm I\kern-.18em H}}
\def\IK{\relax{\rm I\kern-.18em K}}
\def\IL{\relax{\rm I\kern-.18em L}}
\def\IP{\relax{\rm I\kern-.18em P}}
\def\IR{\relax{\rm I\kern-.18em R}}
\def\IZ{\relax\ifmmode\mathchoice
{\hbox{\cmss Z\kern-.4em Z}}{\hbox{\cmss Z\kern-.4em Z}}
{\lower.9pt\hbox{\cmsss Z\kern-.4em Z}}
{\lower1.2pt\hbox{\cmsss Z\kern-.4em Z}}\else{\cmss Z\kern-.4em Z}\fi}
\def\II{\relax{\rm I\kern-.18em I}}

\def\sCC{{\kern 0.27em\vrule height1.45ex width0.03em depth0em
          \kern-0.30em\rm C}}
\def\C{{\mathchoice
  {\sCC}
  {\sCC}
  {\kern 0.225em \vrule height1.05ex width0.025em depth0em \kern-0.25em \rm C}
  {\kern 0.180em \vrule height0.78ex width0.02em depth0em \kern-0.2em \rm C}
        }}
\def\sHH{{\rm I\kern-.16em{}H}}
\def\H{{\mathchoice
  {\sHH}
  {\sHH}
  {\rm I\kern-.13em{}H}
  {\rm I\kern-.13em{}H} }}
\def\sNN{{\rm I\kern-.16em{}N}}
\def\N{{\mathchoice
  {\sNN}
  {\sNN}
  {\rm I\kern-.12em{}N}
  {\rm I\kern-.10em{}N} }}
\def\sPP{{\rm I\kern-.16em{}P}}
\def\P{{\mathchoice
  {\sPP}
  {\sPP}
  {\rm I\kern-.12em{}P}
  {\rm I\kern-.10em{}P} }}
\def\sQQ{{\kern 0.27em \vrule height1.45ex width0.03em depth0em
          \kern-0.30em \rm Q}}
\def\Q{{\mathchoice
        {\sQQ}
        {\sQQ}
  {\kern 0.225em \vrule height1.05ex width0.025em depth0em \kern-0.25em \rm Q}
  {\kern 0.180em \vrule height0.78ex width0.020em depth0em \kern-0.20em \rm Q}
        }}
\def\sRR{{\rm I\kern-0.16em{}R}}
\def\R{{\mathchoice
  {\sRR}
  {\sRR}
  {\rm I\kern-0.12em{}R}
  {\rm I\kern-0.10em{}R} }}
\def\sZZ{{\rm Z\kern-0.32em{}Z}}
\def\Z{{\mathchoice
  {\sZZ}
  {\sZZ} 
  {\rm Z\kern-0.3em{}Z}     
  {\rm Z\kern-0.25em{}Z} }}  
\def\ZZZ{{\rm Z\kern-0.24em{}Z}}
\def\sII{{\rm I\kern-0.16em{}I}}
\def\I{{\mathchoice
  {\sII}
  {\sII}
  {\rm I\kern-0.12em{}I}
  {\rm I\kern-0.10em{}I} }}

\def\inbar{\,\vrule height1.5ex width.4pt depth0pt}
\font\cmss=cmss10 \font\cmsss=cmss10 at 7pt

\def\smiley{\hbox{\large$\bigcirc$\hspace{-0.80em}\raise.2ex
\hbox{$\cdot\cdot$}\kern-.61em\lower.2ex\hbox{\scriptsize$\smile$}}\ }
\def\frowny{\hbox{\large$\bigcirc$\hspace{-0.80em}\raise.2ex
\hbox{$\cdot\cdot$}\kern-.635em\lower.2ex\hbox{\scriptsize$\frown$}}\ }

\def\I{{\rlap{1} \hskip 1.6pt \hbox{1}}}

\makeatletter
\let\hangafter\@hangfrom
\makeatother

\newcommand{\be}{\begin{equation}}
\newcommand{\ee}{\end{equation}}
\newcommand{\bea}{\begin{eqnarray}}
\newcommand{\eea}{\end{eqnarray}}
\newcommand{\bean}{\begin{eqnarray*}}
\newcommand{\eean}{\end{eqnarray*}}
\newcommand{\nn}{\nonumber}
\newcommand{\mN}{\mathcal{N}}
\newcommand{\nf}{\text{\large{n}}^{\hspace{-0.03 cm} \text{\tiny{F}}}}

\preprint{MIT-CTP-3582\\ \\ {\tt hep-th/0412279}}

\title{The Toric Phases of the $Y^{p,q}$ Quivers}

\author{Sergio Benvenuti$^1$, Amihay Hanany$^2$, Pavlos Kazakopoulos$^2$\\

\vspace{0.3 cm}

{1. Scuola Normale Superiore, Pisa,
                   and INFN, Sezione di Pisa, Italy.}\\
~\\
{2. Center for Theoretical Physics,
                   Massachusetts Institute of Technology,\\
                   Cambridge, MA 02139, USA.}\\

~\\

\email{sergio.benvenuti@sns.it,  hanany@mit.edu, noablake@mit.edu}
}
\abstract{We construct all connected toric phases of the recently discovered $Y^{p,q}$ quivers and
show their IR equivalence using Seiberg duality.
We also compute the R and global $U(1)$ charges for a generic
toric phase of $Y^{p,q}$.}
\begin{document}

\section{Introduction}
An interesting class of $\mN = 1$ superconformal gauge theories
can be geometrically engineered placing a stack of D3 branes at
the apex of a Calabi-Yau cone. These  theories are
always \emph{quiver gauge theories}, meaning that all the fields transform in a
two-index representation of the gauge group. They admit a natural
large $N$ limit, and in this limit the gravitational trace anomalies satisfy the relation $c =
a$. The more interesting aspect is that it is possible to take the
near horizon limit \cite{maldacena,MP}: there is a string
dual, provided by Type IIB string theory on \mbox{$AdS_5 \times
X_5$}. $X_5$ is the compact Einstein base of the six-dimensional
cone, which is Calabi-Yau if $X_5$ is \emph{Sasaki-Einstein}. In
order to have a complete description the gauge/string
correspondence it is of course desirable to have the explicit
knowledge of the background, {\it i.e.} of the Sasaki-Einstein metric
on $X_5$.

Until less than a year ago, the explicit metric on $X_5$ was known
only for two homogeneous
Sasaki-Einstein manifolds: $S^5$ and
$T^{1, 1}$. The first case corresponds to $\mN = 4$ SYM. The
second case, the conifold, was analysed in \cite{KW} and
corresponds to a $\mN = 1$ superconformal quiver with gauge group
\mbox{$SU(N) \times SU(N)$}. Of course it is possible to take
orbifolds of these spaces, leading to manifolds with local geometry
of $S^5$ or $T^{1, 1}$. A remarkable development in the field of
Sasakian-Einstein geometry changed this situation: Gauntlett,
Martelli, Sparks and Waldram in \cite{paper1,paper2} found a
countably infinite family of explicit non-homogeneous
five-dimensional Sasaki-Einstein metrics. The corresponding
manifolds are called $Y^{p,q}$, where $q < p$ are positive
integers.


Recently, the dual superconformal field theories were constructed
\cite{Benvenuti:2004dy}. The theories bare the name $Y^{p,q}$ and
they are quiver gauge theories. The precise structure of the
superpotential was found, allowing a comparison
between the global symmetries of the gauge theories and the
isometries of the manifolds. An analogous match was performed for
the baryonic symmetry. As a further non-trivial check of the
gauge/string duality, the volumes of the manifolds and of some
supersymmetric three-cycles were computed in field theory and
matched with geometric results. This was done using the general
field theoretic technique of $a$-maximization, that was also
applied to the known del Pezzo $1$ (corresponding to $Y^{2, 1}$,
\cite{DJ}) and del Pezzo $2$ quivers in \cite{BBC}.


The metric on the $Y^{p,q}$ \cite{paper1,paper2,DJ} in local
form can be written as:
\bea
\label{localmetric}
  \mathrm{d} s_5^2 &=& \frac{1-y}{6}(\mathrm{d}\theta^2+\sin^2\theta\mathrm{d}\phi^2)+\frac{1}{w(y)q(y)}
      \mathrm{d} y^2+\frac{q(y)}{9}(\mathrm{d}\psi-\cos\theta \mathrm{d}\phi)^2 \nonumber \\
      & + &  {w(y)}\left[\mathrm{d}\alpha + f(y) (\mathrm{d}\psi-\cos\theta
      \mathrm{d}\phi)\right]^2
\eea
where
\bea
w(y) & = & \frac{2(b-y^2)}{1-y} \nonumber\\
q(y) & = & \frac{b-3y^2+2y^3}{b-y^2} \nonumber\\
f(y) & = & \frac{b-2y+y^2}{6(b-y^2)}~.
\eea
The coordinate $y$ ranges  between the two smallest roots $y_1,
y_2$ of the cubic $b-3y^2+2y^3$. The parameter $b$ can be
expressed in terms of the positive integers $p$ and $q$:
\beq b =  \frac{1}{2}-\frac{(p^2-3q^2)}{4p^3}\sqrt{4p^2-3q^2}~.
\eeq The topology of the five-dimensional $Y^{p,q}$ spaces is
$S^2\times S^3$. The isometry group is $SO(3)\times U(1)\times
U(1)$ for both $p$ and $q$ odd, and   $U(2)\times U(1)$ otherwise.
This shows up as global symmetry of the quiver gauge theories.  We
will not enter into the details of these metrics, and we refer the
reader to \cite{DJ} for an in-depth exposition.

On the other side of the correspondence one finds the $Y^{p,q}$
quiver gauge theories. These were constructed in
\cite{Benvenuti:2004dy} where it was shown that they can be
obtained from the $Y^{p,p}$ theory.  The five-dimensional
$Y^{p,p}$ space is not smooth, but can be formally added to the
list of the $Y^{p,q}$ spaces, and is the base of the
$\IC^3/\IZ_{2p}$ orbifold. The action of the orbifold group on the
three coordinates of $\IC^3$, $z_i, i=1,2,3$ is given by
$z_i\rightarrow\omega^{a_i}z_i$ with $\omega$ a $2p$--th root of
unity, $\omega^{2p}=1$, and $(a_1,a_2,a_3)=(1,1,-2).$ The dual
gauge theory is easily found. To get the $Y^{p,q}$ theories, one
starts from  $Y^{p,p}$ and applies an iterative procedure $p-q$
times. We will discuss the details of this method in the next
section. At the IR fixed point, one can use Seiberg duality
\cite{Seiberg:1994pq} to find an infinite class of theories that
are inequivalent in the UV but flow to the same conformal fixed
point in the IR. We call these the \emph{phases} of  the $Y^{p,q}$
theories. A finite subclass of these are the so-called \emph{toric
phases}. These theories have the property that all gauge groups in
the quiver have the same rank and every bifundamental field
appears in the superpotential exactly twice: once with a positive
sign and once with a negative sign.  These properties make
explicit the fact that the geometry transverse to the D3 branes is
toric (hence the name).  The IR equivalence of such theories (also
called `toric duality') was discovered in \cite{Feng:2001xr},
interpreted as Seiberg duality in
\cite{Beasley:2001zp,Feng:2001bn} and further elaborated in
\cite{Feng:2002zw,Feng:2002kk}.

The purpose of this note is to construct all the \emph{connected}
toric phases of the  $Y^{p,q}$ quivers. These are the toric phases
that can be reached by applying Seiberg duality on self-dual gauge
groups, {\it i.e.} $SU(N)$ gauge groups with \mbox{$N_f = 2 N_c$}
flavors, whose rank remains the same after the duality. Starting
from a toric phase, one gets another toric phase by dualising a
self-dual node of the quiver. By studying the phases we get  from
these dualisations we will derive a method for constructing all
the connected toric phases of the $Y^{p,q}$ theories as
combinations of different types of `impurities' on the $Y^{p,p}$
quiver. We also demonstrate the agreement between properties of
these quivers and geometric predictions by computing the
R--charges, and show how one can break conformal invariance (while preserving supersymmetry)
by adding fractional branes.

\section{The connected toric phases}

In this section we construct the connected toric phases of the
$Y^{p,q}$ quivers.  As mentioned in the introduction, the term
`connected' means the that we are only considering the toric
phases that can be reached by applying Seiberg duality
 on self-dual gauge groups. We do not have a
general proof that these are all the toric phases, and it is in
principle possible that there are toric phases that can only be
reached by going through non-toric ones. However, our experience
with a number of examples leads us to believe that this is in fact
impossible. For instance, in the case of $3$-block and $4$-block
chiral quivers, the classification of \cite{Benv-Hanany} implies
that all the toric phases are indeed connected. It will be
interesting to find a proof of this. A general property of the
toric phases (when they exist, as is the case for the $Y^{p,q}$
quivers), for any superconformal quiver, is that they are always
`minimal models', or `roots' of the Duality Tree. This can be seen
in the following way. By the definition of the toric phase all the ranks of
the gauge groups are equal. This implies that the `relative number
of flavors' $\nf\equiv \frac{N_f}{N_c}$ is always a positive integer number. For
instance in the models constructed in \cite{Benvenuti:2004dy} one
always find $\nf = 2$ or $\nf = 3$, meaning that there are gauge
groups with $N_f = 2 N_c$ or $N_f = 3 N_c$. Now, if
successive application of Seiberg dualities results in a phase with some $\nf = 1$, 
a problem would occur, since the IR of a gauge group with
$N_f = N_c$ is not superconformal. This would be a contradiction
with the results obtained by Seiberg in \cite{Seiberg:1994pq}. The
conclusion is that for any toric phase all the relative number of
flavors are integer numbers satisfying $\nf \geq 2$. This is
precisely the condition \cite{Benv-Hanany} for a model to be a
root of the Duality Tree, i.e. a (local) minimum for the sum of
the ranks of the gauge groups. In all the models discussed in this
paper, $\nf$ will be equal to $2$, $3$, or $4$.

The $Y^{p,q}$ gauge theories can be built starting from $Y^{p,p}$
through an iterative procedure described in detail in
\cite{Benvenuti:2004dy}.  The $Y^{p,p}$ quiver has a particularly
simple form. It has $2p$ nodes, each representing an $SU(N)$ gauge
group, that can be placed at the vertices of a polygon. If we
number the nodes with an index \mbox{$i, \; i = 1, \ldots , 2 p$}
in a clockwise direction, then between  nodes  $i$ and $i+1$ there
is a double arrow $X_i^{\alpha}, \alpha=1,2$, representing two
bifundamental fields that form a doublet of  the $SU(2)$ global
symmetry and between nodes $i$ and $i-2$ there is a single arrow
$Y_i$ (a singlet of the same $SU(2)$). For example the quiver for
$Y^{44}$ is shown in the upper left corner of \fref{single}.
Following the conventions of \cite{Benvenuti:2004dy}, we denote
the doublets on the outer polygon as $U_i=X_{2i}, V_i=X_{2i+1}$.
In \fref{single} the $U$ fields are colored cyan, the $V$ fields
green and the $Y$ fields  blue. The superpotential for this theory
consists of all possible cubic terms contracted in a fashion that
makes it an invariant of the  $SU(2)$ global symmetry.  It is
written: 
\beq 
W=\sum_{i=1}^{p}\epsilon_{\alpha\beta}(U_i^\alpha
V_{i}^\beta Y_{2i+2}+V_i^\alpha U_{i+1}^\beta Y_{2i+3}). 
\eeq 
The iterative procedure that produces  $Y^{p,q}$ is as follows:
\begin{itemize}
\item Pick an edge of the polygon that has a $V_i$ arrow\footnote{Picking
a $V$ arrow instead of a $U$ is purely a matter of convention,
since $U$ and $V$ are  equivalent in $Y^{p,p}$.}
starting at node $2i+1$, and remove one arrow from the
corresponding doublet to make it a singlet. Call this type of
singlet $Z_i$.

\item Remove the two diagonal singlets, $Y$ that are connected to the two ends of this singlet $Z$.
Since the $V_i$ arrow which is removed starts at node $2i+1$ the
$Y$ fields which are removed are $Y_{2i+2}$ and $Y_{2i+3}$. This
action removes from the superpotential the corresponding two cubic
terms that involve these $Y$ fields.

\item Add a new singlet $Y_{2i+3}$  such that together with
the two doublets at both sides of the singlet $Z_i$,  an oriented
rectangle is formed. Specifically this arrow starts at node $2i+3$
and ends at node $2i$. The new rectangle thus formed contains two
doublets which as before should be contracted to an $SU(2)$
singlet. This term is added to the superpotential.

\end{itemize}

For $Y^{p,q}$ one has to apply the procedure $p-q$ times.  For
example, a phase of $Y^{42}$ is shown at the upper right side of
\fref{single}.  The $Z$ singlets are shown in red.  The added $Y$
singlets are shown in blue. That they have the same color and
notation is justified by the the fact that, as shown in
\cite{Benvenuti:2004dy}, they have the same R-charge and global
$U(1)$ charges as the $Y$ singlets of $Y^{p,p}$.  We will use the
term `impurity' for each 3-step substitution in the $Y^{p,p}$
quiver as above. In this language, $Y^{p,q}$ contains $p-q$
impurities.
An important point is that $Y^{p, p-1}$, and in general $Y^{p,
q}$, is a conformal gauge theory with $c = a$ only at the IR fixed
point.

We must emphasize that what we call IR fixed point is really a manifold
of fixed points, as also discussed in \cite{Ejaz:2004tr}. On the
string theory side, it is possible to modify the background
changing the vev of the axion-dilaton, and giving a vev for the
complex B-field over the $S^2$ (there is precisely one such
possible vev since the second Betti number of the $Y^{p, q}$
manifolds is always $1$). On the gauge theory side this
corresponds, respectively, to a simultaneous rescaling of the
gauge and superpotential couplings, and to a relative change in
the gauge couplings (there is precisely one gauge coupling
deformation since the kernel of the quiver matrix is always two).
This discussion implies that the conformal manifold is at least
two-complex dimensional. It would be nice to see if there are
additional marginal directions, corresponding in the gauge theory
to exactly marginal superpotential deformations and in the
supergravity to continuosly turning on vevs for the other Type IIB
forms (these deformations would probably break the $SU(2)$ global
symmetry).

Also note that these IR fixed points, for finite $q
\neq p$, are not perturbatively accessible. One way to see this is
that there are always \emph{finite} anomalous dimensions for the
bifundamental fields (and so for all  chiral operators), and
this is clearly inconsistent with a fixed point were all the
couplings are infinitesimal. A simple way to see that there are
always finite anomalous dimensions is by noting that in all the
phases of the quivers there are always some gauge groups with 
$\nf = 2$; the numerator of the NSVZ beta function vanishes with
infinitesimal anomalous dimensions only if $\nf = 3$. All Seiberg
dual phases share the same property, since the chiral spectrum is
invariant under Seiberg duality.

\subsection{Seiberg duality moves the impurities}
The above procedure gives toric phases of $Y^{p,q}$.  All nodes
(gauge groups) have rank $N$ and every field enters the
superpotential exactly twice.  However, there is a freedom
involved in this construction, namely the choice of positions for
the impurities.  There are $p$ available positions (the positions
of the $V$ doublets) and  $p-q$ impurities to distribute. The
resulting theories are generally different in the UV.  We will now
show that they are equivalent at the IR fixed point, related by
Seiberg duality. First note that all nodes in  $Y^{p,p}$ have $\nf
= 3$, so none of them is self-dual. Placing the impurities changes
\begin{figure}[ht]
\epsfxsize = 12cm \centerline{\epsfbox{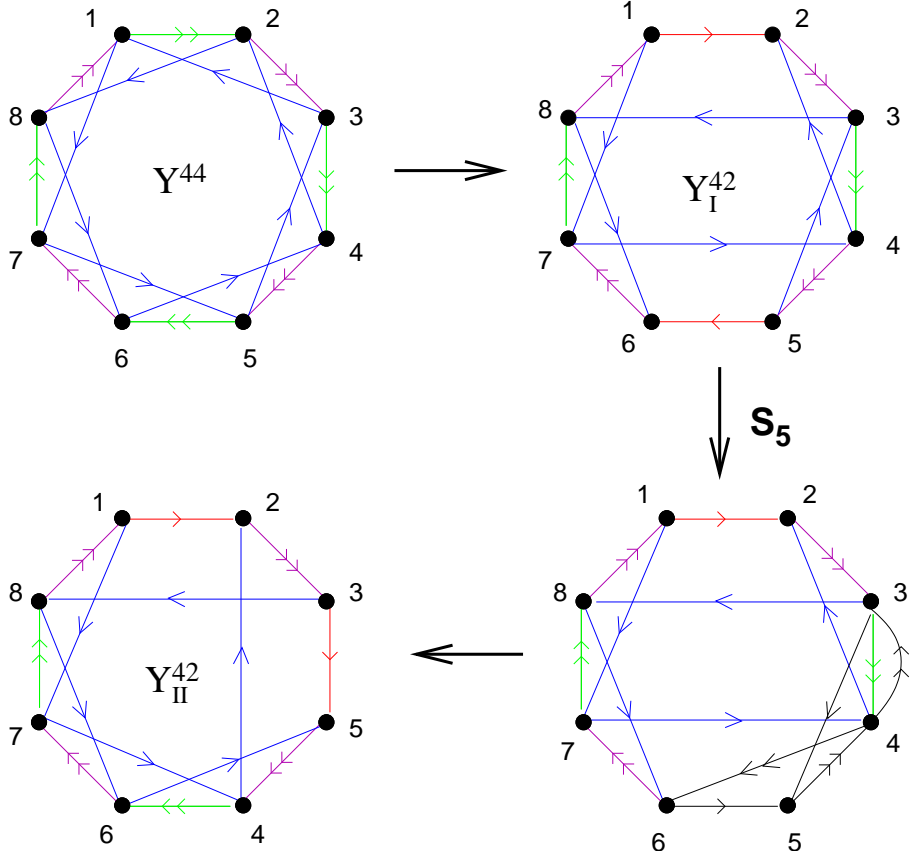}}
\caption{Seiberg duality moves the impurities. The notation $S_5$
means Seiberg duality on node 5.}
  \label{single}
\end{figure}
the relative number of flavors from three to two for the nodes at the ends
of the $Z$ arrows. So the only self-dual nodes in  $Y^{p,q}$ are
the ones at the ends of the $Z$ arrows. Dualising any one of these
nodes will result in a different toric phase.  We illustrate this
using the example of $Y^{42}$. Phase I (the notation is arbitrary)
is shown in the upper right side of \fref{single}. We have four
choices on which node to dualise: Nodes 1, 2, 5, 6 are self-dual.
We choose node 5 and dualise as usual. The new quarks and mesons
are shown in black in the  lower right side of \fref{single}. Note
that the mesons $M_{43}^{\alpha}$ and $M_{46}^{\alpha}$ are
products of a doublet and a singlet of the $SU(2)$ isometry and
thus transform as doublets. The quartic term in the superpotential
involving nodes  4, 5, 6, 7 becomes a cubic term with nodes 4, 6,
7, and another cubic term,
$\epsilon_{\alpha\beta}X_{54}^{\alpha}M_{46}^{\beta}X_{65}$ is
added to the superpotential. The cubic term involving nodes
\mbox{3, 4, 5} becomes a quadratic term
$\epsilon_{\alpha\beta}V_1^{\alpha}M_{43}^{\beta}$ which gives
mass to these fields, so it must be integrated out in the IR
limit.  Integrating out these fields we get a new quartic term
involving nodes \mbox{2, 3, 5, 4.}
 After
eliminating the fields that are integrated out and exchanging
nodes 4 and 5 we obtain the quiver shown in the lower left side of
\fref{single}. But this is exactly what we get from a different
placement of the two impurities. We can describe the effect of
Seiberg duality by saying that the impurity has moved by one step.
This was also shown in \cite{Ejaz:2004tr} and put to good use in
computing  duality cascades for $Y^{p,p-1}$ and $Y^{p,1}$ . It is
easy to see that if we had dualised node six the impurity would
have moved one step in the opposite direction in exactly the same
fashion.  Also, the result of the dualisation  depends only on the
fact that there is no impurity between nodes 3 and 4. The rest of
the quiver goes along for the ride. So dualising one of the nodes
at the ends of a $Z$ field moves the impurity one step in the
direction of the dualised node, as long as there is no impurity
already there.  We have shown that the different phases one gets
from applying the iterative procedure are indeed toric duals. This
fact was briefly mentioned in \cite{Benvenuti:2004dy}.

\subsection{Double impurities}
\begin{figure}[ht]
\epsfxsize = 12cm
\centerline{\epsfbox{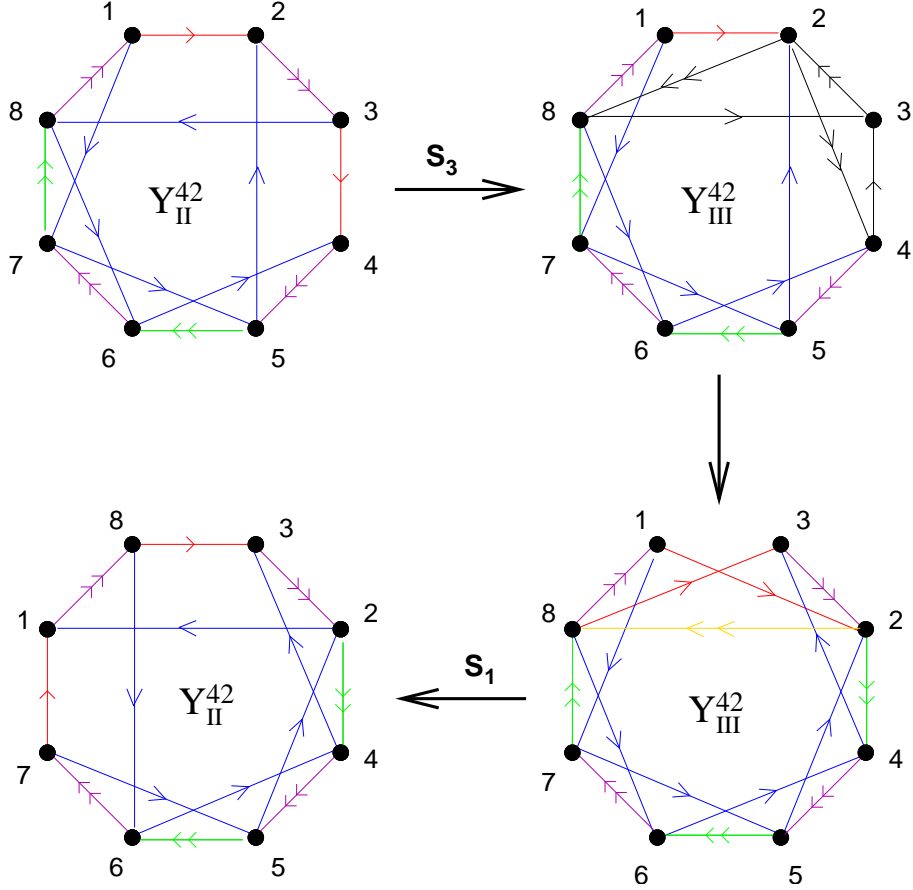}}
\caption{Single impurities can fuse into double impurities.}
  \label{double}
\end{figure}

The next step in constructing the toric phases of $Y^{p,q}$ is to
examine what happens when two impurities `collide'.  We saw before
how Seiberg duality on a self dual-node moves the impurity by one
step. However, when two impurities are adjacent something
different happens. We can illustrate this using  $Y^{42}$ as an
example. It will become clear that the result can be generalized
to any  $Y^{p,q}$ because the duality affects only the vicinity of
the dualised node. We can start from phase II of $Y^{42}$
(\fref{double}).  Nodes 1, 2, 3, 4 are self-dual. Dualising nodes
1 or 4 will move the impurities as before. Dualising nodes 2 or 3
leads to a new phase. We choose to dualise node 3. The new quarks
and mesons are shown in black. The quartic terms associated with
the impurities become cubic terms with nodes 1, 2, 8 and 2, 4, 5,
and new cubic terms
$\epsilon_{\alpha\beta}X_{32}^{\alpha}M_{24}^{\beta}X_{43}$ and
$\epsilon_{\alpha\beta}X_{32}^{\alpha}M_{28}^{\beta}X_{83}$ are
added to the superpotential according to the prescription of
Seiberg duality. After rearranging the nodes we see the new phase
$Y_{III}^{42}$ in \fref{double}. The mesons $M_{28}^{\alpha}$ are
shown in golden because as we will see they have different
R-charges than the fields we have encountered so far. We denote
these fields as $C^{\alpha}$.

This is a new toric phase, different from the ones constructed
from the procedure of \cite{Benvenuti:2004dy}, but equivalent to
these at the IR fixed point.  An interesting thing to note is that
this phase includes only cubic terms in the superpotential (true
only for two impurities) and therefore it is a perturbatively
renormalizable gauge theory. A closer look at this quiver shows
that it actually can be seen formally as a result of applying the
procedure of \cite{Benvenuti:2004dy} \emph{twice} on the same $V$
doublet. We call this a \emph{double impurity}. So applying
Seiberg duality to a self dual node moves the impurity when there
is an `empty slot', but in the case where there is already another
impurity there, the two impurities fuse into a double impurity. It
is clear that the result of this dualisation does not depend on
the rest of the quiver and so it is not specific to $Y_{II}^{42}$.
Two adjacent single impurities can be `fused' in this fashion in
any $Y^{p,q}$. In  $Y_{III}^{42}$, the only self dual nodes are 1
and 3. Dualising node 3 will lead back to $Y_{II}^{42}$, since two
successive dualisations on the same node always give back the same
theory.  In exactly the same way, dualising node 1 will break up
the double impurity into two adjacent single impurities, giving
back the $Y_{II}^{42}$ model.

\begin{figure}[ht]
\epsfxsize = 12cm
\centerline{\epsfbox{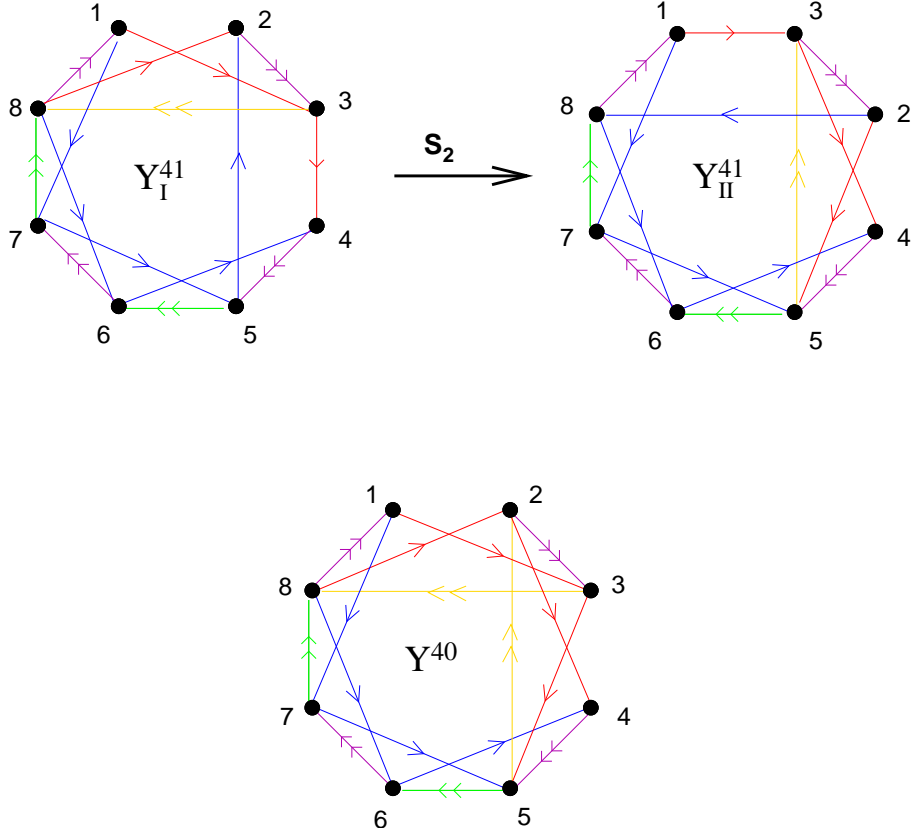}}
\caption{Models with one single and one double impurity and with two double impurities.}
  \label{singledouble}
\end{figure}

A picture is starting to emerge: Single impurities can be moved
around and fused into double impurities, double impurities can be
broken into single impurities and all these models are toric
phases. In this fashion one can think of Seiberg duality as the
`motion of free particles on a circle'. It remains to see what
happens when double impurities `collide' with single impurities or
other double impurities. The answer is that nothing new happens,
and single and double impurities are the only possible
configurations in toric phases.  \fref{singledouble}  illustrates
this.  We see a phase of  $Y^{41}$  with a double impurity next to
a single impurity, labeled $Y_I^{41}$.  Node 3 has $\nf = 3$ and
dualising it will give a non-toric phase. Nodes 1, 2, 4 are self
dual.  We already know that dualising node 1 will separate the
double impurity into two single ones, and give a model with three
single impurities. Dualising 4 will move the single impurity by
one step in a clockwise direction.  Dualising 2 will also break
the double impurity, but the single impurity that is created fuses
with the single impurity next to it to give another double
impurity. We get a different phase with one single and one double
impurity, labeled  $Y_{II}^{41}$ in  \fref{singledouble} (note the
rearrangement of nodes 2 and 3).   A phase of  $Y^{40}$ with two
double impurities is also shown in the figure. The only self-dual
nodes are 1 and 4.  Dualising either one separates the
corresponding double impurity as before.

In all these models, all cubic and quartic gauge invariants in the
quiver enter the superpotential. Each of these terms contains two
$SU(2)$ doublets which are contracted into an  $SU(2)$ singlet.
Single impurities contribute quartic terms, double impurities
cubic terms, and we also have the cubic terms carried over from
$Y^{p,p}$. We can now  state the final result: All connected toric
phases of $Y^{p,q}$ can be constructed by placing $n_1$ single
impurities and $n_2$ double impurities, with $n_1 + 2 n_2 = p-q$,
on $n_1 + n_2$ of the $p$ available positions\footnote{Note that
only the relative positions of the impurities matter.} of the $V$
doublets of  $Y^{p,p}$.  We have also seen how Seiberg duality
connects all these models by moving, fusing and separating the
impurities.  It is worth mentioning that those models that contain
only double impurities have cubic superpotentials and therefore
are renormalizable quantum field theories.  
We note that a double
impurity still 'occupies' $4$ nodes of the quiver, in the sense
that there are $4$ consecutive nodes with $\nf \neq 3$. This
explains why it is impossible to merge together a lot of
impurities, and is consistent with the fact that there are only
single and double impurities.
It is also easy to see
that turning on a non-zero vev for the $p-q$ $Z$ fields in all
these models higgses the quiver to the one for the orbifold $\IC^3
/ \IZ_{p + q}$. We will now proceed to compute the R--charges for
a generic toric phase.

\section{R--charges for a generic toric phase}

We can compute the R--charges of any toric phase using
$a$-maximization \cite{Intriligator:2003jj}. The non-R global
symmetry group in all of the models that we have constructed is
$SU(2)\times U(1)_B \times U(1)_F$. All fields transform in either
singlets or doublets of the $SU(2)$ and are charged under the
global $U(1)$'s.  Since there are two $U(1)$'s with which the R
symmetry can mix there will be two unknowns in the
$a$-maximization. Because of the presence of the $U(1)$-flavor the
R--charges can be irrational (if only baryonic
$U(1)$ symmetries, with vanishing cubic 't Hooft anomalies, are
present, one has to maximize a quadratic function). The trial
R--charge must be anomaly-free (which is equivalent to the
vanishing of the NSVZ beta functions for the $2p$ gauge groups)
and all terms in the superpotential must have R--charge two.

The following assignment satisfies these conditions:
\begin{itemize}
\item The $(p-q)$ singlets $Z$ have R--charge $x$.
\item The $(p+q)$ diagonal singlets $Y$ have R--charge $y$.
\item The $p$ doublets $U$ have R--charge $1-\frac{1}{2}(x+y)$.
\item The $p-(n_1+n_2)$ doublets $V$ have R--charge $1+\frac{1}{2}(x-y)$.
\item The $n_2$ doublets $C$ have R--charge  $1-\frac{1}{2}(x-y)$
\end{itemize}

The quiver structure of the gauge theory automatically implies
that the linear 't Hooft anomaly $\mathrm{tr}R$ vanishes, since it
is given by a weighted sum of the gauge coupling beta functions
\mbox{$\mathrm{tr}R = \sum N_i \beta_i$} \cite{Benv-Hanany}. The
cubic 't Hooft anomaly $\mathrm{tr}R^3$, proportional to the
gravitational central charges $c = a$ \cite{Anselmi:1998am,Anselmi:1998ys}, is given by:
\bea \mathrm{tr}R^3_{\mathrm{trial}}( x, y )&= & 2p +(p-q)(x-1)^3+
(p+q)(y-1)^3-\frac{p}{4}(x+y)^3  +\frac{p-n_1-n_2}{4} (x-y)^3-
\frac{n_2}{4} (x-y)^3\nn\\
&=&2p +(p-q)(x-1)^3+ (p+q)(y-1)^3-\frac{p}{4}(x+y)^3  +\frac{q}{4}
(x-y)^3~. \eea We have used the relation $n_1 + 2 n_2 = p-q$. The
expression for $\mathrm{tr}R^3_{\mathrm{trial}}(x,y)$ is the same
as the one in \cite{Benvenuti:2004dy} and is independent of $n_1$,
$n_2$. As a consequence, the result is the same for all the toric
phases of a given $Y^{p,q}$. This is expected of course, since all
toric phases are related by Seiberg duality. The straightforward
maximization leads to
\bea
x_{max} & = & \frac{1}{3q^2} \left[-4p^2-2pq+3q^2+(2p+q)\sqrt{4p^2-3q^2}\right]\nn\\
y_{max} & = & \frac{1}{3q^2} \left[-4p^2+2pq+3q^2+(2p-q)\sqrt{4p^2-3q^2}\right]~.
\eea
The R--charges and global U(1) charges for the fields are shown in
Table \ref{charges}.
\begin{table}[ht]
\begin{center}
$$\begin{array}{|c|c|c|c|c|}  \hline
 \mathrm{Field}  &\mathrm{number} & R -\mathrm{charge}  & U(1)_B &  U(1)_F \\ \hline\hline

Z    &  p - q     & (- 4 p^2 + 3 q^2 - 2 p q + (2 p + q)\sqrt{4
p^2 - 3 q^2})/3q^2  & p + q &    1    \\\hline

Y &  p + q & (- 4 p^2 + 3 q^2 + 2 p q + (2 p - q)\sqrt{4 p^2 - 3
q^2})/3 q^2 & p - q &   - 1  \\  \hline

U^{\alpha}     &   p & (2 p (2 p - \sqrt{4 p^2 - 3 p^2}))/3 q^2 &
- p   &   0 \\  \hline

V^{\alpha}     &   p-(n_1+n_2)        & ( 3 q - 2 p + \sqrt{4 p^2
- 3 q^2})/3 q   &   q   &   + 1  \\\hline

C^{\alpha} & n_2  &  ( 3 q + 2 p - \sqrt{4 p^2 - 3 q^2})/3 q & - q
& - 1 \\\hline

\end{array}$$
\caption{Charge assignments for the five different types of fields
in the general toric phase of $Y^{p, q}$.} \label{charges}
\end{center}
\end{table}
The R--charges and the central charge computed via field theory
methods match exactly with the geometric data of the volume of
supersymmetric three-cycles and the $Y^{p,q}$ manifolds themselves
\cite{Benvenuti:2004dy,DJ}.

The determination of the baryonic charges leads immediately to the
determination of the vector of the ranks of the gauge groups in
the presence of fractional branes, useful in the study of duality
cascades \cite{Franco:2003ja,Franco:2003ea,Franco:2004jz},
\cite{Ejaz:2004tr}. The reason is that the $U(1)_B$ symmetry is a
linear combination of the $2 p$ decoupled gauge $U(1)$s,
corresponding to one of the two null vectors of the quiver matrix.
It is important that in Table \ref{charges} we chose the
convention that the baryonic charges are always integers numbers.
 
The procedure for changing the ranks, without developing ABJ
anomalies (corresponding to the addition of fractional branes), is
very simple and is as follows.
\begin{itemize}
\item Start with all the ranks of the gauge groups equal to $N$. This
corresponds to the absence of fractional branes.
  \item Pick a node $I$ and change the gauge group from $SU(N)$ to, say, $SU(N + M)$.
  \item Pick an arrow starting from $I$ and arriving
at node $J$. This arrow $I \rightarrow J$ has a well defined
integer baryonic charge $U(1)_B^{I \rightarrow J}$. The rank of
the group at node $J$ is precisely $N + M + U(1)_B^{I \rightarrow
J} M$. For instance, if there is a $U$-field one has \mbox{$N + M
- p M$}, if there is a $Z$-field one has \mbox{$N + M + (p+q) M$}.
  \item Pick an arrow starting from $J$ and arriving at
node $K$. Apply the same
procedure as above with $U(1)_B^{J \rightarrow K}$.
  \item Go on until all nodes are covered. In case there
are only
single-impurities, it is enough to do the full loop of
lenght
$2p$, using the baryonic charges of the doublets $U$ and
the
singlets $Z$.
\end{itemize}
 
It is clear that in this way the new gauge theory, while not
conformal if $M \neq 0$, is still free of ABJ gauge anomalies. Of
course there are two possible freedoms in the previous
construction. First, one can add an "overall" $M$ to the gauge
groups, this is equivalent to a shifting in the number of D3
branes at the singularity. Second, it is possible to rescale $M$,
this is equivalent to a rescaling in the number of wrapped D5
branes (or fractional branes).
 
As check of the procedure, note that after any closed loop one
will always find precisely the initial value. This is due to the
fact that any "mesonic" operator (corresponding to close loops in
the quiver) has vanishing baryonic charge. We note that this
simple procedure is valid for any quiver, also in the case where there are more
than one $U(1)$-baryonic symmetries.

\section{Conclusions}
In this note we have shown how to construct the toric phases of
the newly discovered $Y^{p,q}$ quivers using a combination of
single and double impurity modifications of $Y^{p,p}$. The
impurities move along the circle by each step of Seiberg duality
and have the dynamics of free particles on a circle. There is an
infinity of Seiberg duals for each of the $Y^{p,q}$ theories,
forming a duality tree
\cite{Franco:2003ja,Franco:2003ea,Franco:2004jz} and the toric
phases lie at the roots of this tree. The natural question in this
context is to understand the structure of the full duality tree,
including the non toric phases. It would be nice to understand if
the various phases are classified by the solutions of some
Diophantine equation, as is the case for higher del Pezzo quivers
and for all $3$ and $4$-block chiral models.

Another related problem is the computation of the duality cascades
both from the gauge theory and supergravity perspectives.  Very
significant progress on this has already been made in
\cite{Ejaz:2004tr}.


\section{Acknowledgments}
We would like to thank Sebastian Franco, Dan Freedman, Chris Herzog, and Brian Wecht for useful discussions.
S. B. wishes to acknowledge the kind hospitality of CTP, where part of this work was completed.
The research of  A. H. and P. K. was supported in part by the CTP and LNS of
MIT and the U.S. Department of Energy under cooperative research agreement $\#$ DE--FC02--
94ER40818, and by BSF, an American--Israeli Bi--National Science Foundation. A. H. is also
indebted to a DOE OJI Award.


\bibliographystyle{JHEP}

\end{document}